\documentclass[twocolumn]{aastex7}

\hypersetup{linkcolor=blue,citecolor=blue,filecolor=blue,urlcolor=blue}

\usepackage{epsfig}
\usepackage{epstopdf}  
\usepackage{placeins}
\usepackage{graphicx,color}     
\usepackage{amssymb} 
\usepackage{float}
\usepackage{color}     
\usepackage{url}       
\usepackage{amsmath}
\usepackage{rotating}           
\usepackage{float}            
\usepackage{textcomp}           
\usepackage{dcolumn}
\usepackage{times}
\usepackage{natbib}
\usepackage{placeins}  
\usepackage{xcolor}
\usepackage{hyperref}
\usepackage{tabularx}
\usepackage[english]{babel}
\usepackage{amssymb}
\usepackage{pifont}
\usepackage{hyperref}
\usepackage[utf8]{inputenc}
\usepackage{bm}
\definecolor{ForestGreen}{RGB}{34,139,34}

\begin{document}

\shorttitle{Nanojets in Eruptive and Confined Flares}
\shortauthors{Bura et al.}
\title{Dynamics of Reconnection Nanojets in Eruptive and Confined Solar Flares}

\author[0009-0000-5018-9735]{Annu Bura}
\affiliation{Indian Institute of Astrophysics, Koramangala, Bangalore 560034, India; {\color{blue}{tanmoy.samanta@iiap.res.in}}}
\affiliation{Pondicherry University, R.V. Nagar, Kalapet 605014, Puducherry, India}
\email{annu.bura@iiap.res.in}  

\author[orcid=0000-0001-9035-3245]{Arpit Kumar Shrivastav}
\affiliation{Southwest Research Institute, 1301 Walnut Street Suite 400, Boulder, CO 80302, USA; {\color{blue}{ritesh.patel@swri.org}}}
\email{arpit.shrivastav@swri.org}

\author[orcid=0000-0001-8504-2725]{Ritesh Patel}
\affiliation{Southwest Research Institute, 1301 Walnut Street Suite 400, Boulder, CO 80302, USA; {\color{blue}{ritesh.patel@swri.org}}}
\email{ritesh.patel@swri.org}

\author[orcid=0000-0002-9667-6392]{Tanmoy Samanta} 
\affiliation{Indian Institute of Astrophysics, Koramangala, Bangalore 560034, India; {\color{blue}{tanmoy.samanta@iiap.res.in}}}
\affiliation{Pondicherry University, R.V. Nagar, Kalapet 605014, Puducherry, India}
\email{tanmoy.samanta@iiap.res.in}

\author[orcid=0000-0002-4241-627X]{Sushree S Nayak} 
\affiliation{Departament de F\'isica, Universitat de les Illes Balears, E-07122,
Palma de Mallorca, Spain}
\affiliation{Institute of Applied Computing and Community
Code (IAC3), UIB, E-07122, Palma de Mallorca, Spain}
\affiliation{Indian Institute of Astrophysics, Koramangala, Bangalore 560034, India; {\color{blue}{tanmoy.samanta@iiap.res.in}}}
\email{s.sangeetanayak93@gmail.com}

\author{Ananya Ghosh}
\affiliation{Lady Brabourne College, University of Calcutta, West Bengal, 700017, India}
\affiliation{Aryabhatta Research Institute of Observational Sciences, Nainital, 263002, India}
\email{ananyaghoshanu2003@gmail.com}

\author[0000-0003-4225-8520]{Shanwlee Sow Mondal}
\affiliation{Catholic University of America, DC, Suite 800, Washington, DC 20006, USA}
\affiliation{NASA Goddard Space Flight Center, Greenbelt, MD 27056, USA}
\email{shanwlee.sowmondal@nasa.gov}

\author[orcid=0000-0002-6954-2276]{Vaibhav Pant}
\affiliation{Aryabhatta Research Institute of Observational Sciences, Nainital, 263002, India}
\email{vaibhav.pant@aries.res.in}

\author[orcid=0000-0002-0494-2025]{Daniel B. Seaton}
\affiliation{Southwest Research Institute, 1301 Walnut Street Suite 400, Boulder, CO 80302, USA; {\color{blue}{ritesh.patel@swri.org}}}
\email{daniel.seaton@swri.org}

\begin{abstract}
Recent observations reveal small-scale reconnection-driven plasma ejections, often termed nanojets, triggered by magnetic field interactions at slight misalignment angles. These fast, collimated plasma ejections are $\sim$1.5 Mm long and $\sim$0.5 Mm wide. In this study, we analyze two high-resolution extreme ultraviolet imaging datasets from the Extreme Ultraviolet Imager onboard the Solar Orbiter mission, corresponding to an eruptive (M7.6) and a confined (C1.2) flare, to investigate the dynamics of nanoflare ejections and, for the first time, compare their properties in distinct magnetic environments. We identified 59 nanoflare ejections: 44 in the eruptive flare and 15 in the confined flare event. Our analysis reveals that these events form two distinct classes: confined events exhibit lower speeds (41 – 174\,km s$^{-1}$) and lower kinetic energies (10$^{20}$ -- 10$^{22}$\,erg), placing them closely in or near the picoflare energy regime, while eruptive events show higher speeds (131 – 775\,km s$^{-1}$) and higher kinetic energies (10$^{22}$ -- 10$^{24}$\,erg), falling within the nanoflare regime. Furthermore, magnetic field extrapolations reveal a highly sheared arcade with greater twist and higher magnetic energy density in the eruptive event, compared to the less twisted configuration in the confined event. We infer that this sheared arcade configuration in the eruptive event creates favorable conditions for higher speeds and kinetic energies, unlike the less braided structure in the confined event. Our findings highlight the crucial role of the surrounding magnetic environment in regulating the energetics of nanoflare ejections in the solar atmosphere.
\end{abstract}

\section{Introduction}
    \label{s-intro}
    
The mechanisms responsible for heating the solar corona to temperatures exceeding a million degree Kelvin remain a central question in solar physics. Among the various heating mechanisms proposed, magnetic reconnection occurring at multiple spatial and temporal scales is considered one of the primary candidates \citep{Cirtain2013NaturC, 2017ApJ...843L..20T, 2019Sci...366..890S, 2023Sci...381..867C, 2023ApJ...945...28R, 2024NatAs...8..697B, 2025ApJ...983..144B}. The most prominent reconnection-based heating model is the nanoflare model, originally proposed by \citet{1988ApJ...330..474P}. In this scenario, random continuous convective motions in the photosphere perturb the magnetic footpoints, leading to the formation of increasingly braided and twisted magnetic field lines in the corona (\citealt{1972ApJ...174..499P}; \citeyear{1983ApJ...264..635P}; \citeyear{1983ApJ...264..642P}; \citeyear{1988ApJ...330..474P}). As the angular misalignment between neighboring magnetic strands increases, thin current sheets can form, enabling small-scale reconnection transients \citep{2020ApJ...899...19C, 2021NatAs...5...54A, 2021A&A...656A.141P, 2022ApJ...934..190S, 2022ApJ...938..122P} and flows \citep{Pant2015ApJ}. This process, often termed component magnetic reconnection, is triggered by transverse field components that are misaligned with respect to the dominant guided field line. These reconnection events release energy in the range of nanoflares ($\sim$10$^{24}$ ergs) with the dissipated magnetic energy being converted into thermal energy, kinetic energy, and particle acceleration \citep{Aschwanden2004}.

A recent study utilizing the Interface Region Imaging Spectrograph \citep[IRIS; ][]{2014SoPh..289.2733D} and the Atmospheric Imaging Assembly \citep[AIA;][]{2012SoPh..275...17L} on board the Solar Dynamics Observatory \citep[SDO;][]{2012SoPh..275....3P} has identified a dynamic counterpart to reconnection-driven nanoflares, known as nanojets. \citet{2021NatAs...5...54A} first reported these tiny, collimated bursts which measure approximately 1.5 Mm in length and 0.5 Mm in width, are short-lived (often lasting less than 15 seconds), and exhibit high-velocity plasma flows (100 – 200\,km\,s$^{-1}$). These jets were initially found to appear during coronal rain phenomenon and in newly formed, braided coronal loops in their observation. They usually point radially inward along the curvature of coronal loops. This unidirectional behavior contrasts with the symmetric, bi-directional jets typically observed in standard reconnection scenarios \citep{1992PASJ...44L.173S, 2015Natur.523..437S, 2019ApJ...871..220S}, highlighting a different mechanism linked to the slingshot effect from the magnetically tensed, curved magnetic field lines reconnecting at small angles.

\begin{figure*}[t]
\renewcommand{\thefigure}{1}
\centering
\includegraphics[width=\textwidth]{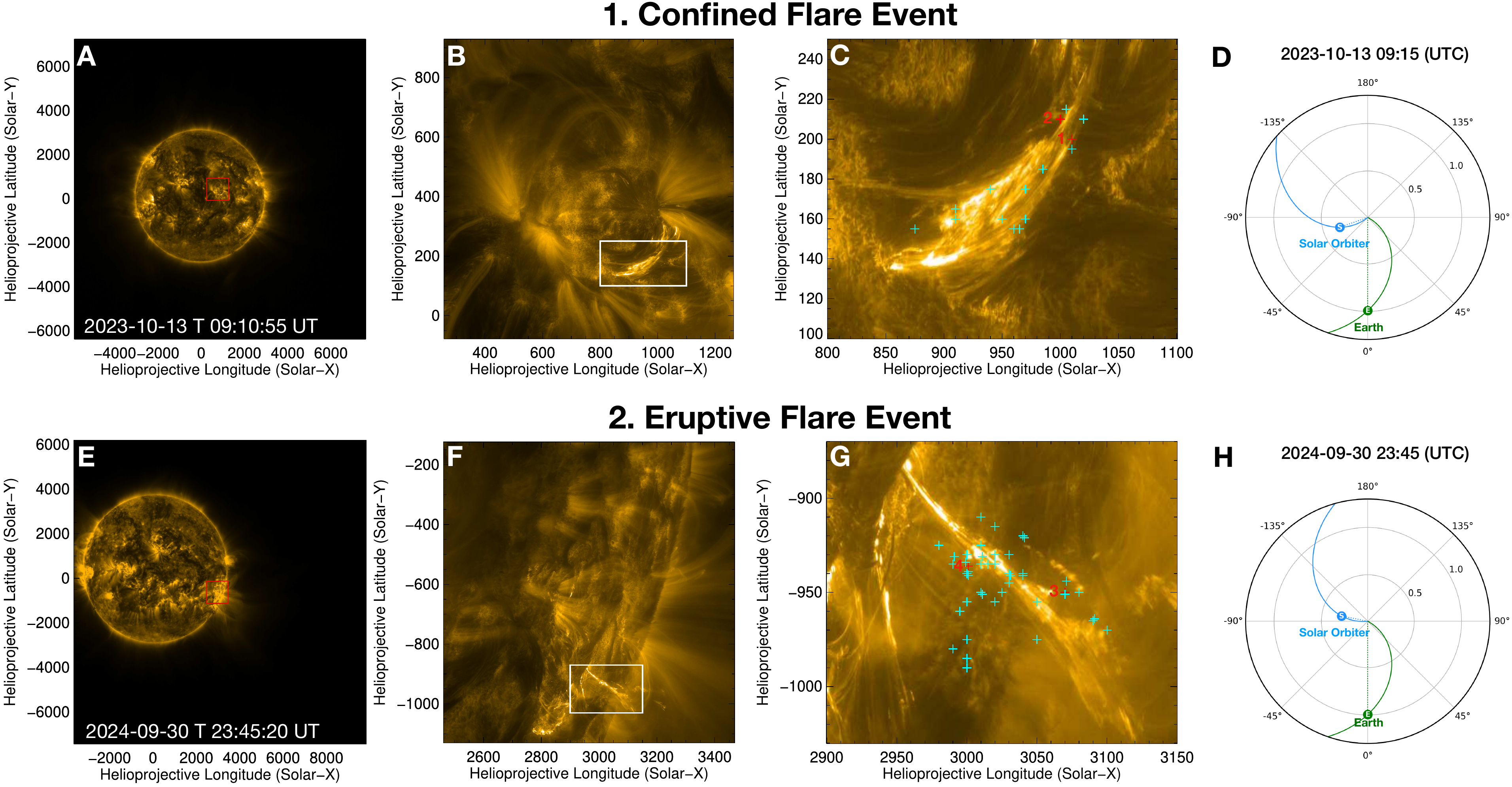}
\caption{Panels (A) and (E) display Full Sun Imager (FSI) images from EUI onboard Solar Orbiter in the 174 \AA\ passband, taken at 09:10:55 UT on October 13, 2023 (confined flare event), and at 23:45:20 UT on September 30, 2024 (eruptive flare event), respectively. The red boxes in panels (A) and (E) indicate the fields of view (FOV) shown in panels (B) and (F). The white boxes in panels (B) and (F) highlight our region of interest, shown in panels (C) and (G). The cyan plus signs indicate the location of nanoflare ejections. The red numbered markers (1 to 4) in panels (C) and (G) denote the locations of the four representative nanoflare ejections in Figure~\ref{fig: fig2}. Panels (D) and (H) illustrate the relative positions of Solar Orbiter and Earth on the corresponding observation dates \citep{2023FrASS...958810G}. An animation of panel (C) (from 09:00 \,UT to 09:30\,UT) and panel (G) (from 23:30\,UT to 23:48\,UT) is available online. The real-time duration of the animation is 16 s.  }
\label{fig: fig1}
\end{figure*}

Using analytical modeling and magnetohydrodynamic simulations, \citet{2021A&A...656A.141P} demonstrated that inward-propagating nanojets are generally more prevalent and energetic than their outward-moving counterparts, primarily due to the dominance of inward-directed magnetic tension forces. However, this asymmetry diminishes as the angle between the reconnecting magnetic field lines increases. 
\citet{2022ApJ...934..190S} further observed nanojets during a blowout jet, also within coronal loops and coronal loops with coronal rain. Their findings suggest that instabilities such as the Kelvin-Helmholtz and Rayleigh-Taylor may facilitate magnetic reconnection in these environments, leading to nanojet formation. In a recent study, \citet{2022ApJ...938..122P} identified nanojets within coronal loops using a High-Resolution Coronal Imager \citep[Hi-C\,2.1;][]{2019SoPh..294..174R} and reported a relatively higher incidence of outward-directed events.

Recent studies suggest that small, blob-like structures - often interpreted as plasmoid - within reconnection current sheets, or within highly dynamic loops during their formation, may be closely linked to the generation of nanojets \citep{2021NatAs...5...54A, 2022ApJ...934..190S, 2023ApJ...943..156K, 2024A&A...687A.190H, 2025ApJ...985...17Z}. Using the Extreme Ultraviolet Imager \citep[EUI;][]{2020A&A...642A...8R} onboard Solar Orbiter \citep{2020A&A...642A...1M}, \citet{2025ApJ...985...17Z} reported bright plasma blobs ejected perpendicular to the loop axis during the relaxation of tangled loops, accompanied by propagating loop brightenings. These signatures likely result from magnetic tension release during unbraiding and localized heating at shifting reconnection sites.

Among the previously discussed works \citep{2021NatAs...5...54A, 2021A&A...656A.141P, 2022ApJ...934..190S, 2022ApJ...938..122P, 2024ApJ...961L..17S, 2024A&A...687A.190H}, nanojets are found to have apparent speeds typically ranging between 50 and 300\,km\,s$^{-1}$, based on time–distance measurements. However, \citet{2025ApJ...985L..12G} recently identified 27 nanojets with average apparent speeds reaching up to 450\,km\,s$^{-1}$. These high speeds are linked to the untwisting of magnetic field lines in an erupting filament and provide new insights into reconnection-driven dynamics in the corona.

In this study, we compare the dynamics of reconnection nanojets in two distinct scenarios: an eruptive and a confined/non-eruptive flare (\citealt{2018SoPh..293...16S}), to explore how the surrounding magnetic environment influences their formation, evolution, and energetics.

\begin{figure*}[t]
\renewcommand{\thefigure}{2}
\centering
\includegraphics[width=\textwidth]{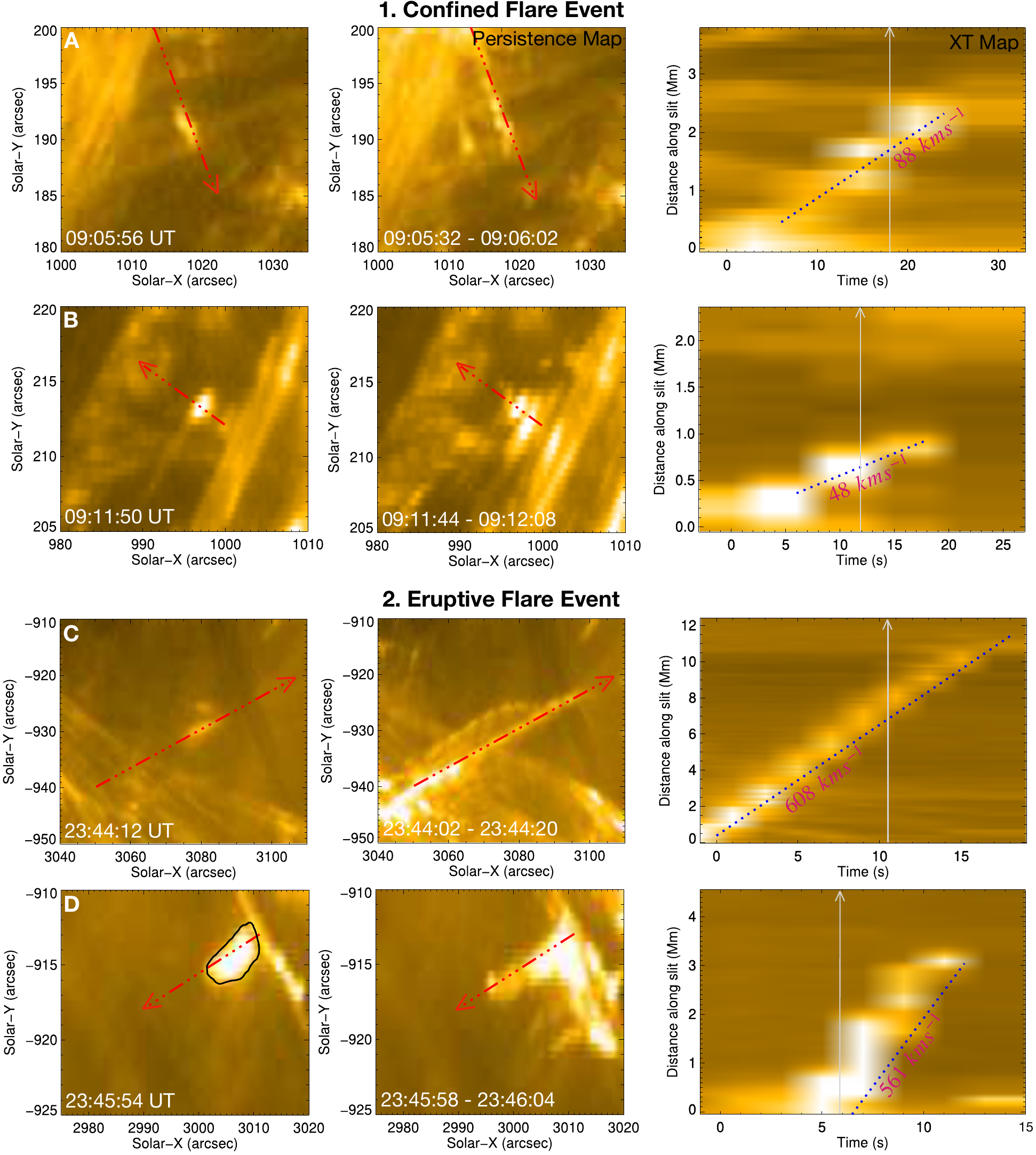}
\caption{Examples of nanoflare ejections observed in both confined and eruptive solar flare events. Panels of (A)- Left: Outward-directed nanoflare eruption during a confined flare, observed in HRI$_{EUI}$ 174\,\AA\ passband at 09:05:56\,UT; Middle: persistence map showing the eruption trajectory; Right: distance–time map along the red artificial slit (arrow indicates slit direction), the blue dotted line traces the bright ridge used to estimate the eruption speed. Panels of (B): Inward-directed nanoflare eruption during a confined flare at 09:11:50\,UT, with similar analysis as in panels of (A). Panels of (C): Outward-directed nanoflare eruption during an eruptive flare at 23:44:12 UT. Panels of (D): Inward-directed nanoflare eruption during an eruptive flare at 23:46:00 UT. The black contour marks a 2$\sigma$ intensity contour used to estimate the length and width of nanoflare ejections.}
\label{fig: fig2}
\end{figure*}

\section{Observations}
    \label{s-observation}
    
We utilized two datasets obtained by High Resolution Imager (HRI) onboard Solar Orbiter on October\,13,\,2023 (hereafter D1), and September\,30,\,2024 (hereafter D2). Specifically, we analyzed HRI$_{\text{EUV}}$ data centered at 174\,\AA. This channel is primarily sensitive to plasma at temperatures around $\sim$1 MK due to contributions from \ion{Fe}{9} (171.1\,\AA) and \ion{Fe}{10} (174.5\,\AA\ and 177.2\,\AA) emission lines.

During the D1 observation, Solar Orbiter was located at a heliocentric distance of $\sim$0.32\,AU (panel D, Figure~\ref{fig: fig1}). HRI$_{\text{EUV}}$ imaged a region of active region (AR) NOAA 13465 (panels A-C, Figure~\ref{fig: fig1}) from 09:00\,UT to 10:07\,UT, while our time of interest is between 09:00\,UT  and 09:30\,UT. A C1.2-class flare of confined nature occurred in this AR, starting at 09:02\,UT and ending at 09:18\,UT. The pixel scale of HRI$_{\text{EUV}}$ is 0.492$''$, corresponding to a pixel size of $\approx$110\,km from this heliocentric distance of Solar Orbiter. The cadence of the observation is 6 seconds.

For the D2 dataset, Solar Orbiter was at a closer distance of $\sim$0.29\,AU (panel H, Figure~\ref{fig: fig1}). HRI$_{\text{EUV}}$ observed a portion of AR NOAA 13842 (panels E-G, Figure~\ref{fig: fig1}) from 23:00\,UT to 23:55\,UT, while our time of interest, during which ejections are visible, is between 23:30\,UT and 23:48\,UT. During this period, HRI$_{\text{EUV}}$ observed at a cadence of 2 seconds with a pixel scale of  $\approx$105\,km. A M7.6 flare that occurred in this AR has recently been studied by \citet{2025ApJ...985L..12G} and R. Patel et al. (under review). The D2 observations also captured an erupting filament exhibiting significant intensity enhancement along with an apparent untwisting motion of the embedded magnetic field lines (\citealt{2025ApJ...985L..12G}). The full width at half maxima of the point spread function of the HRI$_{EUV}$ is about 2 pixels \citep{2021A&A...656L...4B, 2023Sci...381..867C}, corresponding to a spatial resolution of $0.984''$ which is $\sim220$ km for D1 and $\sim210$ km for D2.

We used level 2 datasets of HRI$_{\text{EUV}}$ for both these events. To correct for residual spacecraft jitter in the data, we followed the approach described in \citet{2022A&A...667A.166C}, also employed in earlier HRI$_{\text{EUV}}$ observations \citep{2024Shrivastav, 2025Shrivastav}. Figure~\ref{fig: fig1} provides an overview of the two distinct events investigated in this study: an eruptive (panels A-C) and a confined (panels D-G) flare. These events are selected to examine how the different magnetic environments influence the dynamics of these nanoflare ejections, commonly referred to as $`$nanojets' in earlier studies, but hereafter termed $`$nanoflare ejections' in the context of our analysis and results presented in subsequent sections.

\begin{figure*}[t]
\renewcommand{\thefigure}{3}
\centering
\includegraphics[width=\textwidth]{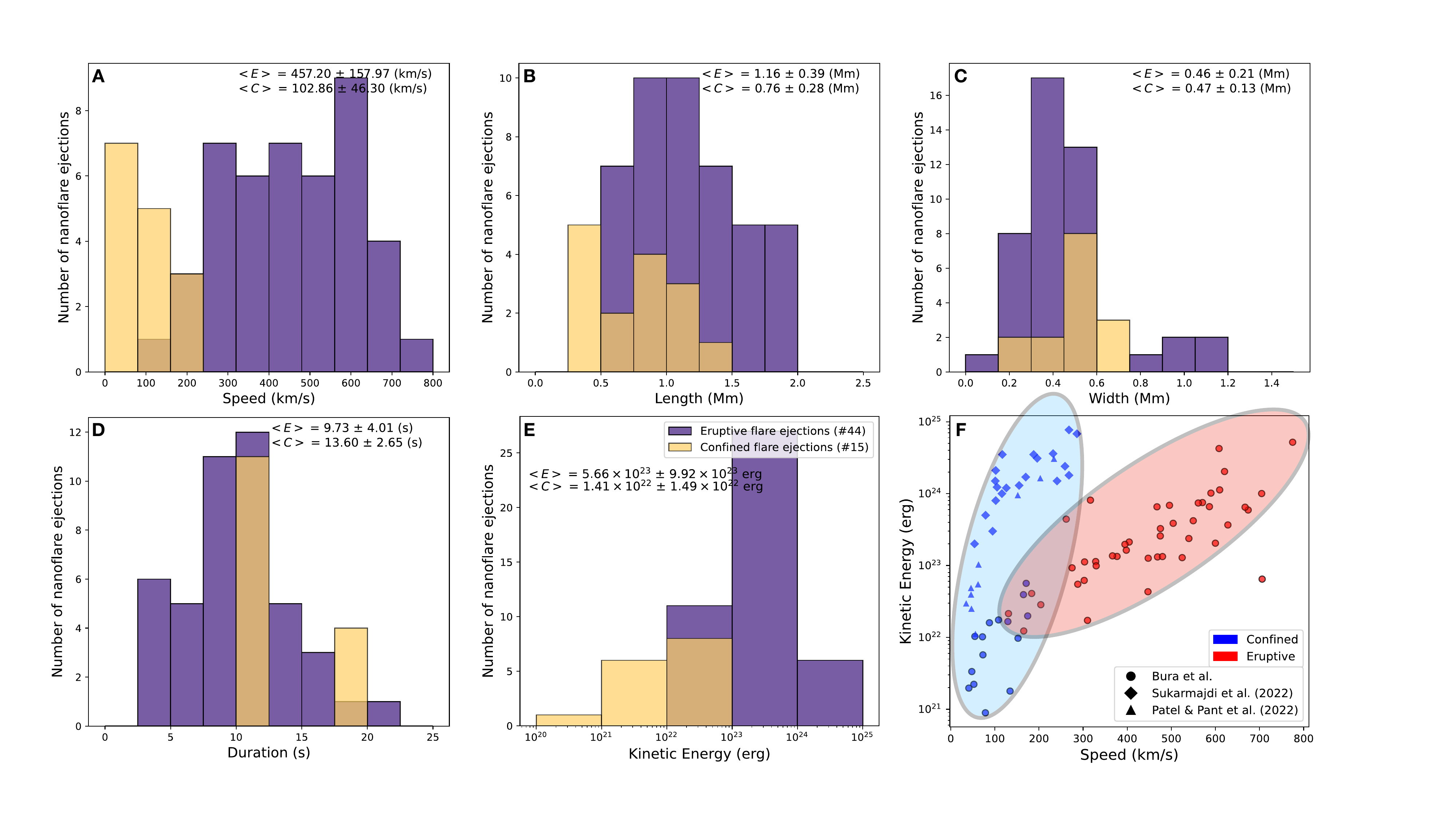}
\caption{Panels (A)-(E): Histograms showing the distribution of speed, length, width, duration, and kinetic energy for 59 nanoflare ejections observed during the D1 and D2 observation periods from Solar Orbiter/HRI, with the number of nanoflare ejections on the y-axis and the respective parameters
on the x-axis. Panel (F) shows the kinetic energy and speed scatter plot for 89 nanoflare ejections, including 59 from our study (depicted as circles with black outlines), 10 from \citet{2022ApJ...938..122P}, and 20 from \citet{2022ApJ...934..190S}, where we can clearly see the bifurcation of the distribution of kinetic energy of nanoflare ejections with speeds: confined ejections (blue-shaded region) occupy the lower speed regime, while eruptive ejections (red-shaded region) exhibit higher speeds.}
\label{fig: fig3}
\end{figure*}

\section{Data Analysis and Results}
    \label{s-analysis}
    
This study investigates nanoflare ejections during two solar flare events, eruptive (M7.6) and confined (C1.2). The eruptive event is distinguished by the presence of an erupting filament, which leads to a dynamic and expansive eruption. In contrast, the confined event is characterized by confined flows, with no large-scale eruption observed. A total of 15 ejections were identified in the braided coronal loops associated with the confined flare, while 44 were observed during the eruptive flare. We selected only those ejections that were visible in at least two frames. The location of these ejections is marked with cyan plus signs in panels (C) and (G) of Figure~\ref{fig: fig1}. For the purpose of our analysis, we considered inward- and outward-directed ejections together, as the primary objective is to compare the properties of nanoflare ejections across different flare environments.

To analyze the temporal evolution and estimate the apparent speeds in the plane-of-sky of the nanoflare ejections, we constructed space–time (XT) diagrams along artificial slits aligned in the direction of the propagation of nanoflare ejections. The left panels in Figure~\ref{fig: fig2} show the artificial slit used for analysis, indicated by a red dashed line and an arrow showing the direction along which the XT diagrams were extracted. The middle panels present the corresponding persistence maps (\citealt{2016ApJ...825...27T}), where each pixel reflects the maximum intensity recorded at that location over a specified time interval, indicated at the bottom of the middle panels. This image processing technique is particularly effective in capturing bright features that evolve both spatially and temporally, such as nanoflare ejections. In the right panels of the same figure, we display these XT diagrams for four representative nanoflare ejections. 

To quantify the apparent speed in a uniform and automated way, we first performed a Gaussian fit along each column of the XT diagrams and determined the peak intensity locations. We then fit these peak positions detected in each column using the \texttt{LINFIT} function in Interactive Data Language (IDL) to determine the slope of the bright ridge, thereby estimating the apparent speed of the eruption. The duration of each ejection was estimated by visually identifying the number of frames in which the feature remained discernible. Based on this analysis, we find that the nanoflare ejections in the confined flare exhibit speeds ranging from 41 to 174\,km\,s$^{-1}$ (mean:~102.86\,km s$^{-1}$) and durations between 12 and 18 seconds (mean:~13.60\,s), whereas those in the eruptive flare show significantly higher speeds ranging from 131 to 775\,km\,s$^{-1}$ (mean:~475.20\,km\,s$^{-1}$) and durations between 4 and 20 seconds (mean:~9.73\,s).

To estimate the length and width of the nanoflare ejections, we used a 2$\sigma$ intensity contour on HRI$_{\text{EUV}}$ 174\,\AA\ images (panel D, Figure~\ref{fig: fig2}); the length was measured along the direction of propagation near the time of maximum extent, and the width was measured perpendicular to it at the center of the nanoflare ejections in the same frame. In the confined flare, the length of ejections is estimated to be between 0.38-1.29\,Mm, with a mean length of 0.76\,Mm, and widths in between 0.18-0.69\,Mm, with a mean width of 0.47\,Mm. In the eruptive flare, lengths varies from 0.53 to 1.95\,Mm, with a mean of 1.16\,Mm, and widths ranges from 0.13 to 1.06\,Mm, with a mean of 0.46\,Mm.

We assumed nanoflare ejections have cylindrical shape, and calculated their volume using \( V = \pi L \left( \frac{w}{2} \right)^2 \), where \( L \) is the length of the eruption and \( w \) is their width, following the method outlined by \citet{2021ApJ...918L..20H}. To estimate the kinetic energy, we used \( E_k = 0.5 n_e m_p v^2 V \) equation \citep{2014ApJ...790L..29T, 2020ApJ...899...19C, 2025ApJ...985L..12G}, where \( n_e \) is the electron number density, \( m_p \) is the proton mass, and \( v \) is the speed of the eruption. The electron density for coronal jets typically lies in the range of $10^{9} - 10^{10}$ cm$^{-3}$ (\citep{2011RAA....11.1229Y, 2017A&A...598A..11M, 2022ApJ...934..190S, 2024ApJ...961L..17S}. In this study, we adopt $10^{9}$ cm$^{-3}$ as a lower-limit estimate to avoid overestimating the energy. Based on these assumptions, the kinetic energy of nanoflare ejections in the confined flare ranges from \( 8.92 \times 10^{20} \,\text{erg} \) to \( 5.64 \times 10^{22} \,\text{erg} \), with a mean value of \( 1.41 \times 10^{22} \,\text{erg} \). For the nanoflare ejections occurring in the eruptive flare, the kinetic energy spans from \( 1.23 \times 10^{22} \,\text{erg} \) to \( 5.19 \times 10^{24} \,\text{erg} \), with a mean value of \( 5.66 \times 10^{23} \,\text{erg} \). 

Tables~\ref{table: table1} and~\ref{table: table2} in appendix~\ref{sec-appendixa} provide a comprehensive summary of the properties of nanoflare ejections observed during the confined and eruptive solar flare, respectively. Figure~\ref{fig: fig3} shows the distributions of apparent speed, length, width, duration, and kinetic energy (panels A-E) of nanoflare ejections in both confined and eruptive solar flare cases. Further, we also compared the speeds by increasing this statistic to 89 by combining findings from \citet{2022ApJ...934..190S} and \citet{2022ApJ...938..122P}. The combined distribution is shown in panel (F) of Figure~\ref{fig: fig3}.

\begin{figure*}[t]
\renewcommand{\thefigure}{4}
\centering
\includegraphics[width=\textwidth]{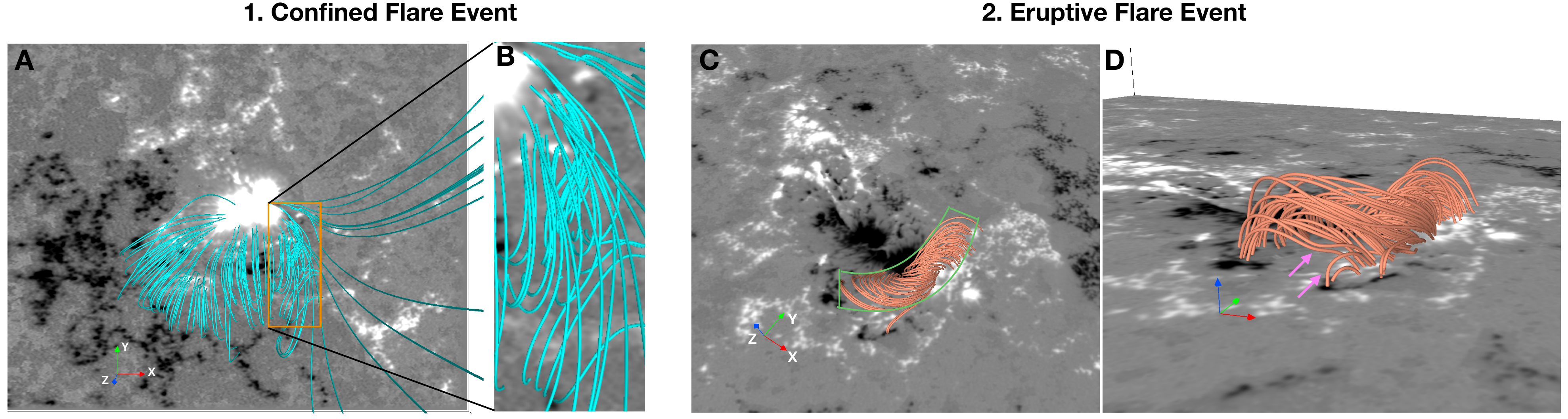}
\caption{Panel (A): NFFF-extrapolated magnetic field lines overlaid on the vertical magnetic field component ($B_{z}$) of AR 13465, shown in grayscale, for the confined flare event at 09:00 UT. Panel (B): Zoomed-in view of the orange box in panel (A), highlighting the region where nanoflare ejections were observed. Panel (C): NFFF-extrapolated field lines plotted over AR 13842 for the eruptive flare event at 23:36 UT, near the site of the filament eruption. Panel (D): Side view of the extrapolated magnetic field configuration shown in panel (C). The orange and green boxes in panels (A) and (C) indicate the regions used for estimating magnetic twist and mean energy density. The x, y, and z directions are indicated at the bottom of panel (A). }
\label{fig: fig4}
\end{figure*}

We further performed the magnetic field extrapolation utilizing the non-force-free-field (NFFF) extrapolation model with the Helioseismic Magnetic Imager \citep[HMI; ][]{2012SoPh..275..207S} magnetogram on board SDO to understand the magnetic field configuration of the two events in our study. The details of the NFFF model can be found in \citet{2007SoPh..240...63B} and \citet{2008ApJ...679..848H}, with some of its application in events on both disk center \citep{2018ApJ...860...96P, 2019ApJ...875...10N, 2024ApJ...975..143N} and off-disk center \citep{2024A&A...691A.198J}. For both events, we have used the SHARP data series (\texttt{hmi.sharp.cea.720s}) available at 12-minutes cadence, which provides vector magnetograms remapped from CCD coordinates to heliographic Cylindrical Equal Area (CEA) projected maps, where projection-related distortions are minimized after de-rotating the patch center \citep{2002A&A...395.1077C, 2014SoPh..289.3549B}. Further, we excluded pixels with noise levels exceeding $\pm250$ G in all three magnetic field components, both while plotting the field lines and estimating the parameters from the extrapolation near the event site, to mitigate the projection effect and/or ill-inverted pixels. For the confined event D1, the region analyzed (panel A, Figure~\ref{fig: fig4}) spans $688\times464\times464$ pixels in the x, y, and z directions (marked at bottom in panel A, Figure~\ref{fig: fig4}), respectively, which corresponds to physical dimensions of $\approx$\,$247\times157\times157$\, Mm.  For eruptive event D2, associated with AR 13842, the region covers $944\times896\times448$ pixels (panel C, Figure~\ref{fig: fig4}), corresponding to about $339\times322\times161$\,Mm. The background of Figure~\ref{fig: fig4} denotes the $B_{z}$ component of the magnetogram for all the panels.

For the confined event, we found a set of twisted magnetic loops near the western side of the AR at 09:00 UT, as shown in panel (A), Figure~\ref{fig: fig4}. The orange box in this panel marks our region of interest, where we observed nanoflare ejections. A closer view of this region is shown in panel (B), revealing several consecutively connected twisted loops. In contrast, for an eruptive event, we found a highly sheared arcade near the site of a filament eruption at 23:36 UT, highlighted in peach color in panel (C). Beneath this arcade, there are additional comparatively low-lying loops with different orientations. These can be seen more clearly from the side view in panel (D), marked by pink arrows.

For a quantitative understanding, we then computed the twist ($T_{w}$) parameter using the code by \citet{liu2016} for both of the events in the region highlighted by orange and green boxes in panels (A) and (C), respectively. We found that the eruptive event has a relatively higher twist ($\approx$ 0.8) than the confined flare event ($\approx$ 0.4). Next, we estimated the non-potential magnetic energy density ($E_{M}=B^2/8\pi$) beyond 1.4\,Mm in height for both events. For confined event, we found mean magnetic energy density $\overline{E_{M}}=$ $8.7\times 10^3$ $G^2$ and for the eruptive event, $\overline{E_{M}}=$ $1.05\times 10^4$ $G^2$. 

\section{Discussion and Conclusion}
    \label{s-discussion}

Magnetic reconnection plays a key role in shaping the solar atmosphere, and hence, having a better understanding of this phenomenon occurring at various scales is important. In this work, we analyze small-scale reconnection-driven nanoflare ejections (popularly known as nanojets) occurring within the braided coronal loop structures of regions undergoing large-scale flares. Our primary objective is to investigate how variations in the surrounding magnetic environment influence the dynamics and energetics of these small-scale ejections. We report on the diversity in nanoflare ejection dynamics associated with solar flares of two different types- an eruptive M7.6-class flare and a confined C1.2-class flare.

We identified and analyzed the physical properties of 15 nanoflare ejections during the confined flare and 44 during the eruptive flare. Our analysis reveals that the nanoflare ejections in the confined flare exhibit an average apparent speed of $\approx$103~km s$^{-1}$ (ranging from 41~km s$^{-1}$ to 174~km s$^{-1}$), whereas those in the eruptive flare reach significantly higher speeds, averaging $\approx$475~km s$^{-1}$ (ranging from 131~km s$^{-1}$ to 775~km s$^{-1}$). The speeds observed in the confined flare are consistent with the standard component reconnection-driven nanojets, which occur across the magnetic field lines, reported earlier by \citet{2021NatAs...5...54A, 2022ApJ...934..190S, 2022ApJ...938..122P, 2024ApJ...961L..17S} and picoflare jets ($\sim$100~km s$^{-1}$) that are directed along the magnetic field lines, recently identified by \citet{2023Sci...381..867C}. In contrast, the considerably higher speeds observed in the eruptive flare case were rarely seen in earlier studies. Although \citet{2025ApJ...985L..12G} reported high-speed reconnection nanojets based on a subset of 27 events from the same eruptive flare event, our study analyzes 44 such ejections, identifying a broader sample and providing a more detailed characterization of the dynamics involved.

We found that the nanoflare ejections in the confined flare have an average lifetime of $\sim$14\,s, while those in the eruptive flare last for $\sim$10\,s. In contrast, earlier studies have reported durations of up to $\sim$80\,s \citep{2022ApJ...938..122P}. This difference likely arises because previous studies predominantly focused on nanojets associated with either coronal loops or coronal rain, which were non-eruptive in nature.
Although the lengths and widths of these jets are consistent with previously reported values \citep{2021NatAs...5...54A, 2022ApJ...934..190S,2022ApJ...938..122P}, their significantly shorter durations distinguish them from earlier observations, indicating a potentially different impact of the surroundings leading to this difference. We also used persistence maps \citep{2016ApJ...825...27T} to enhance the visibility of transient and bright nanoflare ejections, which helped us to trace their individual trajectory. The kinetic energy versus speed scatter plot (Panel F, Figure \ref{fig: fig3}), based on 89 ejections (59 from our study, 10 from \citealp{2022ApJ...938..122P}, and 20 from \citealp{2022ApJ...934..190S}), reveals a clear bifurcation in the distribution of speed for nanoflare ejections associated with confined (blue-shaded region) and eruptive events (red-shaded region). We found that confined ejections have lower speeds, while eruptive ejections exhibit higher speeds.

\begin{figure}[ht]
\centering
\includegraphics[width=\columnwidth]{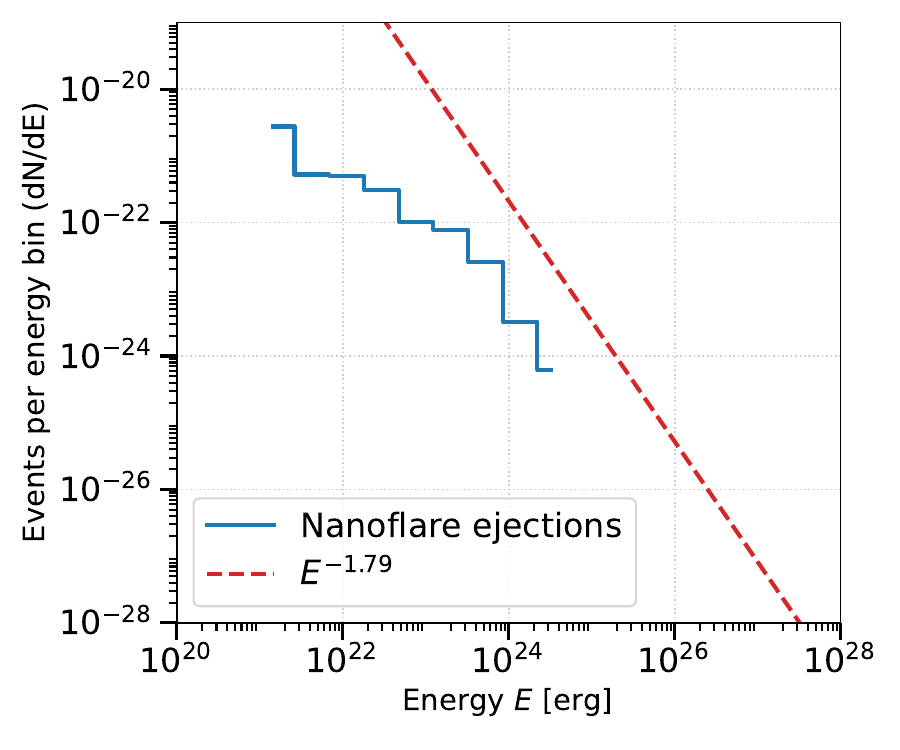}
\caption{The blue histogram shows the distribution of kinetic energies of the 59 nanoflare ejections identified in this study. The red dashed line indicates a reference power-law distribution with a slope of $-1.79$, based on the statistical analysis of EUV flares by \citet{Aschwanden2000ApJ}.} 
\label{fig:powerlaw}
\end{figure}

We further estimated the kinetic energy of the nanoflare ejections and identified two distinct energy regimes. The nanoflare ejections in the confined flare exhibit kinetic energies ranging from 10$^{20}$ - 10$^{22}$\,erg placing them likely within the picoflare energy regime ($10^{21}$\,erg) comparable to energetics of picoflare jets reported by \citet{2023Sci...381..867C}. In contrast, the nanoflare ejections in the eruptive flare have kinetic energies ranging from 10$^{22}$ - 10$^{24}$\,erg, which corresponds to the nanoflare energy ($10^{24}$\,erg) regime \citep{2021NatAs...5...54A}. If a higher density of $10^{10}$ cm$^{-3}$ was assumed, the estimated energy values would increase by one order of magnitude, but the distinction between the two classes of confined and eruptive events would clearly persist, with energies in or near the picoflare and nanoflare regimes, respectively. Based on our analysis of 59 nanoflare ejections, we determined that nanoflare ejections in eruptive flare events exhibit higher speeds and greater energy compared to confined events. 

The blue curve in Figure \ref{fig:powerlaw} represents the distribution of kinetic energies of the 59 combined nanoflare ejections identified in this study. The x-axis denotes the flare energies ($E$), while the y-axis shows the occurrence frequency, expressed as the number of events per unit energy interval ($dN/dE$). The distribution is computed using logarithmic binning with uniform spacing in $log_{10}(E)$, rather than linear energy intervals \citep{1971SoPh...16..152D, 1991SoPh..133..357H, Aschwanden2000ApJ, 2013PASJ...65...49S}. The red dashed line indicates the occurrence frequency of EUV flares from a previous study of \cite{Aschwanden2000ApJ} -- following a power law distribution,

\begin{equation}
\frac{dN}{dE} \propto E^{-\alpha}
\end{equation}
with a slope, $\alpha = 1.79$. The details of binning methods and power-law fitting procedure are provided in appendix~\ref{sec-appendixb}. Comparison between the distributions indicates close alignment in the higher energy range, with deviation in the lower energy range. This deviation can be explained by the following possible reasons. (1) The instrument resolution can limit the detection of nanoflare ejections, thereby limiting the statistics. (2) The observational limitations, such as lower cadence in the confined event dataset relative to the eruptive event dataset and different emission levels in both datasets, might have resulted in the low-brightness and low-speed nanoflare ejections not being detected, further limiting the statistics. (3) Projection effects might also contribute to this deviation due to the underestimation of the measured speed in both events, where a part of the measured speed could lie along the line of sight. (4) The deviation may reflect the presence of different underlying physical mechanisms driving nanoflare ejections and EUV flares. 

The various energy ranges in the two cases hint towards the fact that the braiding reconnection in the confined case is relatively weaker than the eruptive flare case. This is also reflected in the flare X-ray class of the two scenarios, with a stronger flare associated with the eruptive event (M7.6), giving rise to more energetic ejections. Additionally, results from the NFFF extrapolation revealed that the confined event contains twisted magnetic field lines (panel B, Figure~\ref{fig: fig4}). In such configurations, magnetic reconnection can be triggered by misalignments between adjacent field lines, as suggested by \citet{2021NatAs...5...54A}. In contrast, the eruptive flare event exhibits a highly sheared arcade at the site of the filament eruption. With sufficient photospheric shearing or converging motions, this sheared arcade can evolve into a coherent flux rope, potentially leading to an eruption \citep{moore2001, kumar2016, zhao2017}. Moreover, the presence of cross-oriented loops beneath the arcade may promote magnetic reconnection and further support the eruption process. We also found that the eruptive flare event has a higher magnetic twist and a greater mean magnetic energy density than the confined event. The higher twist may lead to larger misalignment between field lines, while the higher magnetic energy density provides more free energy, together enhancing the likelihood and strength of magnetic reconnection. These combined factors likely contribute to the production of more energetic nanoflare ejections in the eruptive event. It is worth noting that these extrapolations were performed prior to the peak of each event; the amount of twist may be even greater during the peak phase, as evident from the animation available online.

Furthermore, in the eruptive event, a more magnetically stressed coronal loop containing a filament was present, which underwent rapid evolution. This is evident from the continuous appearance of new magnetic strands in nearly every observational frame, indicating fast and ongoing magnetic restructuring \citep{2021NatAs...5...54A, 2022A&A...667A.166C,Cirtain2013NaturC}. As the filament underwent apparent untwisting, it likely increased the misalignment between adjacent magnetic field lines, leading to more efficient energy release and faster nanoflare ejections. Further, the increasingly open magnetic field configuration during the eruption may have further reduced resistance to plasma outflows, facilitating higher ejection speeds compared to the confined flare. These high eruption speeds suggest that a greater portion of magnetic energy is released and converted into kinetic energy, indicating a more effective energy dissipation process \citep{chitta2025magneticavalanchecentralengine, 2025ApJ...985L..12G}.  Moreover, the AR and overlying sheared arcade in the eruptive event appear to have undergone a gradual accumulation of magnetic stress over a long period, thereby storing more magnetic energy \citep{1983ApJ...264..635P, 1988ApJ...330..474P, Reid2018AandA}. This not only leads to the large-scale eruption, but is also responsible for providing more energy to their nanoflare ejection counterparts at small scales, depicting energy cascade through magnetic reconnection at various scales \citep{Barta2011ApJ, chitta2025magneticavalanchecentralengine}.

In contrast, the confined flare, which was a weaker C1.2-class event, did not show any large-scale eruption. The magnetic field in this case appeared less dynamic and more resistant to large-scale reconfiguration, with reduced braiding compared to the eruptive flare. In this environment, energy release likely occurred more gradually, with nanoflare ejections driven by the slower relaxation or $`$slingshot' effect of unbraided loops \citep{2021NatAs...5...54A}. These differences indicate a weaker reconnection process and a smaller buildup of magnetic energy in the confined region. This is further supported by the study of \citet{2018SoPh..293...16S}, who found significantly smaller changes in the horizontal magnetic field and Lorentz force during confined flares. Together, these factors suggest that the weaker magnetic restructuring in confined flares might lead to less energetic nanoflare ejections. 

While the underlying mechanism remains consistent for both cases, the efficiency and scale of energy release differ depending on the magnetic environment. The eruptive event, characterized by rapidly evolving and more stressed magnetic fields, provides conditions for high-speed and kinetic energy nanoflare ejections. In contrast, the confined event, with its less misaligned and less complex magnetic structure, results in weaker ejections. With the increase in statistics in the future using high-resolution observations, such transients will provide a new window to probe the very small-scale reconnection dynamics. This will help us understand if the nanoflares and picoflares follow a power law or deviate from this, as suggested by some studies \citep{Ryan2016AnA}. These findings underscore the significant role that magnetic topology and dynamic processes play in shaping the energy released during small-scale solar events.

\begin{acknowledgments}
Solar Orbiter is a space mission with international collaboration between ESA and NASA, operated by ESA. The EUI instrument was built by CSL, IAS, MPS, MSSL/UCL, PMOD/WRC, ROB, LCF/IO with funding from the Belgian Federal Science Policy Oﬃce (BELSPO/PRODEX PEA 4000134088); the Centre National d’Etudes Spatiales (CNES); the UK Space Agency (UKSA); the
Bundesministerium für Wirtschaft und Energie (BMWi) through the Deutsches Zentrum für Luft- und Raumfahrt (DLR); and the Swiss Space Oﬃce (SSO). R.P. and D.B.S. are supported by NASA Solar Orbiter Guest Investigator Grant Number: 80NSSC24K1243. S.S.N. acknowledges the support from the I+D+i project PID2023-147708NB-I00 funded by
MICIU/AEI/10.13039/501100011033/  and  by  FEDER, EU. V.P. acknowledges the support received from the ISRO RESPOND project (RES-PRL-2022-018). 
\end{acknowledgments}

\vspace{5mm}
\facilities{Solar Orbiter (FSI and HRI),
            SDO (HMI)}
\software{astropy \citep{astropy2013, astropy2018},  
          sunpy \citep{sunpy_community2020}, 
          IDL \citep{idl2000}
          }

\appendix 

\section{Appendix A}
\label{sec-appendixa}

\FloatBarrier 
\begin{deluxetable*}{rlllrrr}
\tablenum{A1}
\tablecaption{Properties of nanoflare ejections during a confined solar flare event on October\,13,\,2023. \label{tab:nanojets}}
\tablewidth{0pt}
\label{table: table1}
\tablehead{
\colhead{S.No.} & \colhead{Start Time (UT)} & \colhead{Speed (km s$^{-1}$)} &
\colhead{Length (Mm)} & \colhead{Width (Mm)} & \colhead{Duration (s)} & 
\colhead{Kinetic Energy (erg)} 
}
\startdata
1  & 09:04:08 & 134 & 0.45 & 0.18 & 12 & $1.78 \times 10^{21}$  \\
2  & 09:04:26 & 153 & 0.45 & 0.38 & 12 & $9.72 \times 10^{21}$  \\
3  & 09:04:56 & 79  & 0.38 & 0.24 & 12 & $8.92 \times 10^{20}$  \\
4  & 09:05:02 & 108 & 0.79 & 0.53 & 18 & $1.75 \times 10^{22}$  \\
5  & 09:05:40 & 88  & 1.08 & 0.54 & 18 & $1.59 \times 10^{22}$  \\
6  & 09:05:44 & 41  & 0.81 & 0.47 & 18 & $1.97 \times 10^{21}$  \\
7  & 09:06:14 & 174 & 0.49 & 0.45 & 12 & $1.98 \times 10^{22}$  \\
8 & 09:08:50 & 72  & 0.76 & 0.63 & 18 & $1.02 \times 10^{22}$  \\
9 & 09:09:14 & 130 & 0.70 & 0.46 & 12 & $1.65 \times 10^{22}$  \\
10 & 09:11:50 & 55  & 1.29 & 0.64 & 12 & $1.03 \times 10^{22}$  \\
11 & 09:11:50 & 48  & 0.47 & 0.69 & 12 & $3.34 \times 10^{21}$\\
12 & 09:14:32 & 52  & 0.59 & 0.46 & 12 & $2.22 \times 10^{21}$  \\
13 & 09:15:08 & 165 & 0.99 & 0.47 & 12  & $3.93 \times 10^{22}$  \\
14 & 09:15:56 & 171 & 1.18 & 0.50 & 12 & $5.64 \times 10^{22}$  \\
15 & 09:16:08 & 73  & 1.03 & 0.39 & 12 & $5.71 \times 10^{21}$  \\
\enddata
\end{deluxetable*}

\begin{deluxetable*}{rlllrrrrr}
\tablenum{A2}
\tablecaption{Properties of nanoflare ejections during an eruptive solar flare event on September\,30,\,2024. \label{tab:nanojets2}}
\tablewidth{0pt}
\label{table: table2}
\tablehead{
\colhead{S.No.} & \colhead{Start Time (UT)} & \colhead{Speed (km s$^{-1}$)} &
\colhead{Length (Mm)} & \colhead{Width (Mm)} & \colhead{Duration (s)} & \colhead{Kinetic Energy (erg)} 
}
\startdata
16 & 23:39:52 & 448 & 0.91 & 0.32 & 4 & $1.26 \times 10^{23}$ &  \\
17 & 23:39:56 & 540 & 1.10 & 0.34 & 4 & $2.38 \times 10^{23}$ &  \\
18 & 23:39:58 & 469 & 0.55 & 0.41 & 8 & $1.31 \times 10^{23}$ & \\
19 & 23:40:16 & 586 & 1.02 & 0.53 & 18 & $6.58 \times 10^{23}$ \\
20 & 23:40:16 & 674 & 1.29 & 0.39 & 12 & $5.92 \times 10^{23}$ \\
21 & 23:40:26 & 165 & 0.82 & 0.28 & 10 & $1.23 \times 10^{22}$  \\
22 & 23:40:34 & 525 & 0.95 & 0.27 & 10 & $1.29 \times 10^{23}$  \\
23 & 23:40:46 & 404 & 1.52 & 0.36 & 8 & $2.12 \times 10^{23}$  \\
24 & 23:41:27 & 705 & 1.13 & 0.52 & 8 & $1.00 \times 10^{24}$ \\
25 & 23:41:27 & 706 & 1.13 & 0.13 & 8 & $6.46 \times 10^{22}$ \\
26 & 23:41:37 & 667 & 1.79 & 0.35 & 16 & $6.47 \times 10^{23}$ \\
27 & 23:43:45 & 447 & 0.64 & 0.23 & 14 & $4.32 \times 10^{22}$  \\
28 & 23:43:53 & 496 & 1.62 & 0.51 & 8 & $6.88 \times 10^{23}$  \\
29 & 23:43:54 & 475 & 1.34 & 0.40 & 6 & $3.26 \times 10^{23}$ \\
30 & 23:44:02 & 608 & 1.95 & 0.94 & 20 & $4.24 \times 10^{24}$ \\
31 & 23:44:10 & 262 & 0.87 & 1.05 & 8 & $4.41 \times 10^{23}$  \\
32 & 23:44:12 & 303 & 1.25 & 0.28 & 4 & $6.20 \times 10^{22}$ \\
33 & 23:44:22 & 183 & 1.27 & 0.38 & 6 & $4.08 \times 10^{22}$  \\
34 & 23:44:28 & 504 & 1.05 & 0.47 & 12 & $3.86 \times 10^{23}$  \\
35 & 23:44:32 & 600 & 0.98 & 0.29 & 4 & $2.04 \times 10^{23}$  \\
36 & 23:44:38 & 621 & 1.38 & 0.76 & 16 & $2.03 \times 10^{24}$\\
37 & 23:44:38 & 571 & 1.69 & 0.46 & 10 & $7.51 \times 10^{23}$  \\
38 & 23:44:48 & 377 & 1.84 & 0.28 & 14 & $1.34 \times 10^{23}$  \\
39 & 23:44:48 & 610 & 1.87 & 0.49 & 16 & $1.13 \times 10^{24}$  \\
40 & 23:44:48 & 550 & 1.39 & 0.39 & 8 & $4.19 \times 10^{23}$  \\
41 & 23:45:04 & 275 & 0.53 & 0.59 & 14 & $9.28 \times 10^{22}$ \\
42 & 23:45:04 & 329 & 1.02 & 0.39 & 6 & $1.13 \times 10^{23}$  \\
43 & 23:45:14 & 204 & 0.78 & 0.36 & 12 & $2.85 \times 10^{22}$  \\
44 & 23:45:14 & 367 & 0.97 & 0.39 & 6 & $1.36 \times 10^{23}$  \\
45 & 23:45:16 & 310 & 0.55 & 0.22 & 10 & $1.72 \times 10^{22}$  \\
46 & 23:45:28 & 303 & 0.72 & 0.51 & 12 & $1.12 \times 10^{23}$  \\
47 & 23:45:30 & 475 & 0.73 & 0.48 & 10 & $2.58 \times 10^{23}$ \\
48 & 23:45:42 & 317 & 1.22 & 1.01 & 8 & $8.11 \times 10^{23}$  \\
49 & 23:45:48 & 775 & 1.19 & 1.05 & 10 & $5.19 \times 10^{24}$  \\
50 & 23:45:54 & 561 & 1.44 & 0.49 & 8 & $7.41 \times 10^{23}$  \\
51 & 23:46:00 & 590 & 1.78 & 0.50 & 8 & $1.02 \times 10^{24}$  \\
52 & 23:46:24 & 131 & 1.19 & 0.40 & 6 & $2.14 \times 10^{22}$  \\
53 & 23:46:28 & 330 & 1.54 & 0.41 & 4 & $9.89 \times 10^{22}$  \\
54 & 23:46:42 & 398 & 0.93 & 0.39 & 12 & $1.63 \times 10^{23}$ \\
55 & 23:46:42 & 628 & 0.92 & 0.52 & 8 & $3.66 \times 10^{23}$  \\
56 & 23:46:46 & 468 & 1.67 & 0.59 & 14 & $6.55 \times 10^{23}$  \\
57 & 23:46:48 & 395 & 0.55 & 0.55 & 14 & $1.97 \times 10^{23}$  \\
58 & 23:47:01 & 288 & 0.94 & 0.34 & 10 & $5.99 \times 10^{22}$  \\
59 & 23:47:31 & 480 & 1.03 & 0.29 & 4 & $1.33 \times 10^{23}$  \\
\enddata
\end{deluxetable*}
\clearpage 

\section{Appendix B}
\label{sec-appendixb}

To construct the plot in Figure \ref{fig:powerlaw}, which illustrates the differential energy distribution of solar flares, we adopted the methodology outlined in \cite{Aschwanden2000ApJ, Aschwanden2015ApJ}. In this approach, the differential energy distribution ($dN/dE$) provides the number of events per energy bin ($dE$) and is known to follow: $\mathrm{d}N/\mathrm{d}E \propto E^{-\alpha}$. Hence, the total energy ($E_{tot}$) released by flares in the energy range $[E_{min}, E_{max}]$, is given by:

\begin{equation}
    E_{tot} = \int_{E_{min}}^{E_{max}} \frac{\mathrm{d}N}{\mathrm{d}E} E \,\mathrm{d}E,
\end{equation}
where \( E \) denote the flare energies. It should be noted that the quantity $dN/dE$, representing the number of flares per energy bin, has appeared in various forms in the literature. For example, it is denoted as $N(E)$ in \citet[][and references therein]{Aschwanden2000ApJ, Aschwanden2004, Aschwanden2025arXiv}, while earlier studies have used notations like $N$ \citep{Drake1971SoPh} or $N(P)$ \citep{Dennis1985SoPh} to describe the same distribution.

The differential energy distributions are most intuitively visualized on a logarithmic scale. To construct them, flare events are grouped into logarithmically spaced energy bins to account for the broad range in flare energies. For each energy bin $[E_i, E_{i+1}]$, we compute the number of flare events $N(E_i)$ falling within that bin. The differential energy distribution, $\mathrm{d}N/\mathrm{d}E_i$, is then obtained by dividing the event count by the bin width ($dE_i$). While the energy bins are evenly spaced in $log_{10}E$, the division is performed using actual linear width of the bin ($dE_i = E_{i+1}-E_i$), which is non-uniform in linear energy space.

An alternative way to obtain this similar plot is to convert the parameters in both the X and Y axes to logarithmic values as shown in Figure 1a in \citet{Hudson1991SoPhH}. In this case, $log_{10}(dN/dE)$ vs $log_{10}(E)$ is plotted where both the axes are spaced in linear scale.
In both approaches, the resulting power-law distribution appears as a straight line with a slope of $-\alpha$. The power-law index $\alpha$ is determined by fitting a linear regression to the log-log data over the energy range where the distribution exhibits a clear power-law behavior, often excluding the low-energy end where detection biases may cause incompleteness. This approach enables quantitative comparison of flare energy distributions.

\end{document}